\documentclass{jnmp}

\usepackage{amsmath}
\usepackage{graphicx}

\JNMPnumberwithin{equation}{section}

\begin{document}

\renewcommand{\evenhead}{J~Harnad}
\renewcommand{\oddhead}{Equations for Fredholm Determinants Appearing in Random Matrices}

\thispagestyle{empty}

\FirstPageHead{9}{4}{2002}{\pageref{harnad-firstpage}--\pageref{harnad-lastpage}}{Article}

\copyrightnote{2002}{J~Harnad}

\Name{On the Bilinear Equations for Fredholm Determinants
Appearing in Random Matrices}
\label{harnad-firstpage}

\Author{J~HARNAD}

\Address{Department of Mathematics and Statistics, Concordia University,\\
7141 Sherbrooke W., Montr\'eal, Qu\'e., Canada H4B 1R6, {\rm and} \\
Centre de recherches math\'ematiques, Universit\'e de Montr\'eal,\\
C.~P.~6128, succ. centre ville, Montr\'eal, Qu\'e., Canada H3C 3J7\\
E-mail: harnad@crm.umontreal.ca}

\Date{Received May 11, 2002; Revised May 29, 2002; Accepted
June 12, 2002}

\begin{abstract}
\noindent
It is shown how the bilinear differential equations satisfied by Fredholm
determinants of integral operators appearing as spectral distribution functions
for random matrices may be deduced from the associated systems of nonautonomous
Hamiltonian equations satisfied by auxiliary canonical phase space variables
introduced by Tracy and Widom. The essential step is to recast the latter as
isomonodromic deformation equations for families of rational covariant derivative
operators on the Riemann sphere and interpret the Fredholm determinants as
isomonodromic $\tau$-functions.
\end{abstract}

\section{Differential equations for Fredholm determinants\\ in random matrices}

   In the theory of random matrices, it is known that in suitably defined double
scaling limits the generating functions for spectral distributions are given by
Fredholm determinants of certain integral operators \cite{M,TW1,TW2,TW3}. For example, in
the universality class of the Gaussian Unitary Ensemble (GUE), in the bulk of the
spectrum, the probability of having exactly $\{m_1, \ldots, m_n\}$ scaled
eigenvalues in the sequence of disjoint intervals
$\{([a_1,a_2], \ldots, [a_{2n-1},a_{2n}]\}$ is
\begin{equation}
E(m_1,\ldots,m_n) = \frac{(-1)^{\bar m}}{m_1!\cdots m_n!}
\frac{\partial^{\bar m} \tau^S}{\partial\lambda_1^{m_1}\cdots\partial\lambda_{n}^{m_n}}
\Bigr\vert_{\lambda_1=\cdots=\lambda_m=1}, \qquad \bar m=\sum_{j} m_j,
\end{equation}
where $\tau^S$ is the Fredholm determinant
\begin{equation}
\tau^S:=\det\left(1-\hat{K}^S\right)
\label{sineFredholmdet}
\end{equation}
of the integral operator $\hat{K_s}:L^2({\mathbb R},{\mathbb C})\to L^2({\mathbb R},{\mathbb C})$
 with the sine kernel
\begin{equation}
\left(\hat{K}^S v\right)(x) =\sum_{j=1}^n\lambda_j\int_{a_{2j-1}}^{a_{2j}}
\frac{\sin(\pi(x-y))}{\pi(x-y)}\, v(y)dy.
\end{equation}
Rescaling  at the (soft) edge of the spectrum, the corresponding  quantity is given
by the Fredholm determinant
\begin{equation}
\tau^A:=\det\left(1-\hat{K}^A\right)
\label{AiryFredholmdet}
\end{equation}
of the operator with the Airy kernel \cite{TW2}
\begin{equation}
\left(\hat{K}^A v\right)(x)
=\sum_{j=1}^n\lambda_j\int_{a_{2j-1}}^{a_{2j}}
\frac{Ai(x))Ai'(y)-Ai(y)Ai'(x)}{x-y}\,
v(y)dy,
\end{equation}
where $Ai(x)$ is the Airy function. If the measure is taken to be the
one associated  with either the Laguerre or Jacobi orthogonal polynomials,
rescaling at the (hard) edge leads to the Fredholm determinant
\begin{equation}
\tau_\alpha^B:=\det\left(1-\hat{K}_\alpha^B\right)
\label{BesselFredholmdet}
\end{equation}
of the operator with Bessel kernel~\cite{F, TW3}
\begin{equation}
\left(\hat{K}_\alpha^B v\right)(x) =\sum_{j=1}^n\lambda_j\int_{a_{2j-1}}^{a_{2j}}
\frac{J_\alpha(\sqrt{x})\sqrt{y}J_\alpha'\left(\sqrt{y}\right)-J_\alpha
\left(\sqrt{y}\right)\sqrt{x}J_\alpha'\left(\sqrt{x}\right)}{2(x-y)}\, v(y)dy,
\end{equation}
where $J_\alpha(x)$ is the Bessel function with index $\alpha$.

  It was shown by Tracy and Widom \cite{TW1,TW2,TW3}, extending earlier results of
the Kyoto school~\cite{JMMS}, that all these Fredholm determinants can be computed by
quadratures in terms of solutions of certain associated nonautonomous Hamiltonian
systems in which the end points $\{a_j\}$ play the r\^ole of  multi-time
deformation variables. Moreover, these Fredholm  determinants may be interpreted as
isomonodromic $\tau$-functions~\cite{HTW, P, HI, BD} in the sense of \cite{JMU, JM}.

  More recently, Adler, Shiota and van Moerbeke~\cite{ASV1, ASV2} have shown that
the Fredholm determinants $\tau^A$, $\tau^B_\alpha$ satisfy hierarchies of bilinear
differential equations with respect to the endpoint parameters. These follow
from combining Virasoro constraints satisfied by certain associated KP $\tau$-functions
with the bilinear equations they also satisfy with repect to the KP flow parameters
$\{t_1, t_2, \ldots\}$, evaluated at the zero values of these parameters. The approach
of~\cite{ASV1, ASV2} was based on the application of vertex operators, integrated over
the intervals $\{[a_{2j-1},a_{2j}]\}$,  to suitable ``vacuum'' KP $\tau$-functions,
effecting thereby a continuous version of Darboux transformations, yielding new KP
$\tau$-functions, such that the  Fredholm determinant equals the ratio of the two.

For the Airy kernel, the first equation in this hierarchy may be expressed as
\begin{equation}
{\mathcal D}_0^4F^A - 4{\mathcal D}_1{\mathcal D}_0 F^A + 2
{\mathcal D}_0F^A + 6\left({\mathcal D}_0^2F^A\right)^2 = 0,
\label{ASVAiry}
\end{equation}
where
\begin{equation}
F^A:= \ln\tau^A
\end{equation}
and
\begin{equation}
{\mathcal D}_m:=\sum_{j=1}^{2n}a_j^m\frac{\partial}{\partial a_j},
\qquad m\in {\mathbb N},
\end{equation}
while for the Bessel kernel, it is
\begin{gather}
{\mathcal D}_1^4F_\alpha^B -2{\mathcal D}_1^4 F_\alpha^B +
\left(1-\alpha^2\right){\mathcal D}_1^2F_\alpha^B + {\mathcal
D}_2{\mathcal D}_1F_\alpha^B \nonumber\\ \phantom{{\mathcal
D}_1^4F_\alpha^B} {}-\frac 12{\mathcal D}_2F_\alpha^B -
4\left({\mathcal D}_1F_\alpha^B\right) \left({\mathcal
D}_1^2F_\alpha^B\right) + 6\left({\mathcal
D}_1^2F_\alpha^B\right)^2 = 0, \label{ASVBessel}
\end{gather}
with
\begin{equation}
F_\alpha^B:= \ln\tau_\alpha^B.
\end{equation}
No analogous equations were derived for the sine kernel, although in
the special case where the intervals $[a_{2j-1},a_{2j}]$ are chosen symmetrically about
the origin, the Fredholm determinant $\tau^S$ may be expressed \cite{M, TW3} as a
product $\tau_{\frac 12}^B\tau_{-\frac 12}^B$ of two Bessel kernel determinants.

In the case of a single interval, it is easy to see that equations \eqref{ASVAiry} and
\eqref{ASVBessel} just give the $\tau$-function form of the Painlev\'e equations $P_{II}$
and $P_V$, respectively, to which the Tracy--Widom systems reduce in the case of
the Airy and Bessel kernels. It seems reasonable to expect that analogous results
hold for the general case, involving an arbitrary number of intervals. The purpose of
this work is to show how the hierarchies of equations derived in~\cite{ASV1, ASV2} can
in fact be deduced directly from the Tracy--Widom Hamiltonian systems for both the Airy
and Bessel cases, and to also apply this approach to the sine kernel case. The main
step is to recognize that the Hamiltonian systems imply isomonodromic deformation
equations for associated families of rational covariant derivative operators on the
Riemann sphere. It is known \cite{JMU, JM} that such isomonodromic deformations give
rise to bilinear equations for indexed sets of isomonodromic $\tau$-functions related by
Schlesinger transformations. The fact that for the systems associated with
the Airy and Bessel kernels such equations may be written in terms of a single scalar
$\tau$-function is due to the presence of a pair of conserved quantities,
allowing the elimination of the additional variables  by fixing the level
sets of these invariants. In the sine kernel case this is not possible, and the
associated bilinear equations therefore involve coupled systems for $\tau^S$ together
with a~pair of additional variables $(\tau_+^S,\tau_-^S)$.

  In Section 2, equations  \eqref{ASVAiry} and \eqref{ASVBessel} are first derived directly from the
Hamiltonian systems of \cite{TW2,TW3}.  In Section~3, it is shown how the isomonodromic
deformation equations following from the associated Hamiltonian systems may be used to
derive the full hierarchy of $\tau$-function equations for all these cases.
In section 4, these results are related to the rational classical $R$-matrix approach to
isomonodromic and isospectral systems developed in~\cite{AHP, H}.

\section{Deduction of $\boldsymbol{\tau}$-function equations\\
 from the Hamiltonian systems}

   To establish notation, following \cite{TW1,TW2,TW3}, we define the quantities:
\begin{subequations}\label{xjdef}
\begin{gather}
x_{2j}:= 2i\sqrt{\lambda_j}({\mathbb I}-\hat{K})^{-1}\phi (a_{2j}),
\qquad x_{2j+1}:= 2\sqrt{\lambda_j}({\mathbb I}-\hat{K})^{-1}\phi (a_{2j+1}), \label{xjdef.a} \\
y_{2j} :=i\sqrt{\lambda_j}({\mathbb I}-\hat{K})^{-1}\psi (a_{2j}),
\qquad
y_{2j+1}:=\sqrt{\lambda_j}({\mathbb I}-\hat{K})^{-1}\psi (a_{2j+1}),  \label{yjdef.b}\\
x_0 :=2\sum_{j=1}^n\lambda_j\int_{a_{2j-1}}^{a_{2j}}\phi(x)({\mathbb I}-\hat{K})^{-1}\psi (x)dx,
\label{xzerodef.c}
\end{gather}
\begin{gather}
 y_0 :=\sum_{j=1}^n \lambda_j\int_{a_{2j-1}}^{a_{2j}}\phi(x)({\mathbb I}-\hat{K})^{-1}\phi (x)dx,
\label{yzerodef.d}
\end{gather}
\end{subequations}
where, for the case of the sine kernel $\hat{K}=\hat{K}^S$,
\begin{equation}
\phi(x):= \frac{\sin(\pi x)}{\pi}, \qquad \psi(x): =\cos(\pi x).
\label{xySinedef}
\end{equation}
while for the  Airy kernel $\hat{K}=\hat{K}^A$,
\begin{equation}
\phi(x):= Ai(x), \qquad \psi(x): =\frac{d Ai(x)}{dx}, \label{xydef}
\end{equation}
and for the Bessel kernel $\hat{K}=\hat{K}_\alpha^B$,
\begin{equation}
\phi(x):= J_\alpha(\sqrt{x}), \qquad \psi(x): = x\frac{dJ_\alpha\left(\sqrt{x}\right)}{dx}.
\end{equation}
(An odd number of variables may also occur if we set one of the $a_j$'s equal
to some fixed constant, say $0$ or $\infty$, and eliminate the corresponding
pair $(q_j, p_j)$.) As shown in \cite{TW1,TW2,TW3}, the logarithmic derivatives of the
associated Fredholm determinants are given by:
\begin{equation}
G^S_j:=\frac{\partial F^S}{\partial a_j} =  \frac{\pi^2}{4} x_j^2 + y_j^2  -\frac 14
\sum_{k=1\atop k\neq j}^n \frac{(x_jy_k-y_jx_k)^2}{a_j-a_k}
\label{GSjdef}
\end{equation}
for the sine kernel,
\begin{equation}
G^A_j:=\frac{\partial F^A}{\partial a_j} =
y_j^2 +\frac 14 (x_0 -a_j)x_j^2 -y_0x_jy_j
-\frac 14 \sum_{k=1\atop k\neq j}^n \frac{(x_jy_k-y_jx_k)^2}{a_j-a_k}
\label{GAjdef}
\end{equation}
for the Airy kernel, and
\begin{gather}
a_jG^B_{\alpha,j}:= a_j\frac{\partial F^B_\alpha}{\partial a_j} =
y_j^2 -\frac{1}{16} \left(\alpha^2 -a_j + x_0\right)x_j^2 \nonumber\\
\phantom{a_jG^B_{\alpha,j}:=a_j\frac{\partial F^B_\alpha}{\partial a_j} = } {}+\frac 14 y_0 x_j y_j
-\frac 14 \sum_{k=1\atop k\neq j}^n \frac{a_k(x_jy_k-y_jx_k)^2}{a_j-a_k}
\label{GBjdef}
\end{gather}
for the Bessel kernel.

For use in what follows, we also define the quantities
\begin{equation}
R^S_m:={\mathcal D}_m F^S = \sum_{j=1}^{2n}a_j^mG^S_j, \qquad m\in {\mathbb N}
\label{RSmdef}
\end{equation}
for the sine kernel case,
\begin{equation}
R^A_m:={\mathcal D}_m F^A = \sum_{j=1}^{2n}a_j^mG^A_j, \qquad m\in {\mathbb N}
\label{RAmdef}
\end{equation}
for the Airy case and
\begin{equation}
R^B_{\alpha,m}:={\mathcal D}_m F^B_{\alpha} = \sum_{j=1}^{2n}a_j^m G^B_{\alpha,j},
\qquad m\in {\mathbb N}\label{RBmdef}
\end{equation}
for the Bessel case. For all three cases, we define the following sequence of
bilinear forms
\begin{equation}
P_m :=\sum_{j=1}^{2n}a_j^my_j^2, \qquad
Q_m :=\sum_{j=1}^{2n}a_j^mx_j^2, \qquad
S_m :=\sum_{j=1}^{2n}a_j^mx_j y_j, \qquad m\in {\mathbb N}.  \label{PQS}
\end{equation}

  As explained below, the $\{G^A_j\}$'s and $\{G^B_{\alpha,j}\}$'s may
be viewed as sets of Poisson commu\-ting, nonautonomous
Hamiltonians on an auxiliary phase space with canonical
coordinates $\{x_0,y_0,x_j, y_j\}$, such that the quantities
defined in~\eqref{xjdef} satisfy the corresponding systems of
Hamiltonian equations. These equations will then be shown to imply
equations~\eqref{ASVAiry} and~\eqref{ASVBessel}.

\subsection{The Airy kernel system}

  The system of dynamical equations for this case is given~\cite{TW2} by
\begin{subequations}
\label{AiryHameq}
\begin{gather}
\frac{\partial x_j}{\partial a_k} =-\frac{1}{2} \frac{(x_j y_k-y_jx_k)x_k}{a_j-a_k},
\qquad  j\neq k, \label{AiryHameqa.a}\\
\frac{\partial y_j}{\partial a_k} =-\frac{1}{2} \frac{(x_j y_k-y_jx_k)y_k}{a_j-a_k},
 \qquad j\neq k, \label{AiryHameqb.b}\\
\frac{\partial x_j}{\partial a_j} =\frac{1}{2}\sum_{k=1\atop k\neq j}^n
\frac{(x_j y_k-y_jx_k)x_k}{a_j-a_k} +2y_j -y_0x_j, \label{AiryHameqc.c}\\
\frac{\partial y_j}{\partial a_j}=\frac{1}{2}\sum_{k=1\atop k\neq j}^n
\frac{(x_j y_k-y_jx_k)y_k}{a_j-a_k}+\frac{1}{2}(a_j-x_0)x_j + y_0x_jy_j,
\label{AiryHameqd.d}\\
\frac{\partial x_0 }{\partial a_j} = -x_j y_j,  \qquad
\frac{\partial y_0}{\partial a_j} =-\frac{1}{4}x_j^2. \label{AiryHameqe.e}
\end{gather}
\end{subequations}
Viewing the $a_j$'s as multi-time parameters, and the quantities
$\{x_0,y_0,x_j,y_j\}$ as cano\-ni\-cal coordinates, this is a compatible system of
nonautonomous Hamiltonian equations generated by the Poisson commuting Hamiltonians
$\{G_j^A\}$ defined in~\eqref{GAjdef}. There is an additional functionally independent
Hamiltonian, defined by
\begin{equation}
G^A_0 := y_0^2 -x_0 -\frac{1}{4}Q_0,  \label{GAzero}
\end{equation}
which also Poisson commutes with all the $G_j^A$'s. Since $G_0^A$ is not explicitly
dependent on the parameters $\{a_j\}$, it follows that it is a conserved
quantity. Since all the quantities $\{x_0,y_0,x_j,y_j\}$  defined in
\eqref{xjdef} vanish in the limit $\{a_j \to \infty, \ \forall \; j\}$, the
invariant
$G^A_0$ must vanish on this particular solution. Therefore we may express $x_0$ in terms
of the other variables as
\begin{equation}
x_0=y_0^2-\frac{1}{4}Q_0. \label{xzeroAiry}
\end{equation}

  The quantity $R_0^A$ defined in~\eqref{RAmdef} will just be denoted
\begin{equation}
R:=R^A_0= \sum_{j=1}^{2n}G^A_j = P_0 -\frac{1}{4}Q_1 +\frac{1}{4}y_0^2 Q_0
-y_0S_0 -\frac{1}{16}Q_0^2,  \label{RAeval}
\end{equation}
where \eqref{xzeroAiry} has been used. In terms of $R$, equation \eqref{ASVAiry} becomes
\begin{equation}
{\mathcal D}_0^3 R -4{\mathcal D}_1 R + 2 R + 6 ({\mathcal D}_0 R)^2 = 0.  \label{ASVAiryR}
\end{equation}
It follows from the Poisson commutativity of the Hamiltonians $\{G_j^A\}_{j=1,\ldots, 2n}$
that their Hamiltonian vector fields applied as derivations to $R$ give zero, and hence
along any integral surface of eqs.~\eqref{AiryHameq}, the derivatives of
$R$ with respect to the $a_j$'s are just given by its {\it explicit} dependence on
these parameters. This just comes from the $Q_1$ term in expression~\eqref{RAeval},
and therefore we have
\begin{equation}
\frac{\partial R}{\partial a_j} = -\frac{1}{4}x_j^2 \label{DRAaj}
\end{equation}
Comparing with~\eqref{AiryHameqe.e}, this implies that
\begin{equation}
G^A_\infty := y_0 - R  \label{GAinfy}
\end{equation}
is a second conserved quantity.   Since in the limit $\{a_j\to \infty, \ \forall \; j\}$,
both $y_0$ and $R$ vanish, $G^A_\infty$ must vanish for all values of the parameters,
and therefore the invariant relation
\begin{equation}
y_0=R  \label{yzeroAiry}
\end{equation}
is satisfied by this solution. Applying the operators ${\mathcal D}_0$, ${\mathcal D}_1$ to $R$, it
follows from~\eqref{DRAaj} that
\begin{subequations}
\label{DDzeroR}
\begin{gather}
{\mathcal D}_0 R = -\frac{1}{4} Q_0,  \label{DDzeroR.a}\\
{\mathcal D}_1 R = -\frac{1}{4} Q_1.  \label{DDoneR.b}
\end{gather}
\end{subequations}
Eqs.~\eqref{AiryHameq} also imply that application of ${\mathcal D}_0$ to
$\{Q_0, S_0, x_0, y_0, Q_1\}$ gives
\begin{subequations}\label{PSzeroAderiv}
\begin{gather}
{\mathcal D}_0 Q_0 =4S_0 - 2y_0 Q_0, \qquad
{\mathcal D}_0 S_0 =\frac{1}{2}Q_1 -\frac{1}{2}x_0Q_0 +2P_0, \label{PSzeroAderiv.a}\\
{\mathcal D}_0 x_0 = - S_0, \qquad {\mathcal D}_0 y_0 =-\frac{1}{4}Q_0,  \label{xyAzeroderiv.b}\\
{\mathcal D}_0 Q_1 =Q_0 +4S_1-2y_0 Q_1.  \label{Qonederiv.c}
\end{gather}
\end{subequations}
 Further application of ${\mathcal D}_0$ and ${\mathcal D}_1$, using \eqref{DDzeroR.a},
\eqref{PSzeroAderiv} and \eqref{xzeroAiry}, therefore gives
\begin{subequations}\label{RAtwo}
\begin{gather}
{\mathcal D}_0^2 R = \frac{1}{2}y_0 Q_0 -S_0,  \label{RAtwo.a}\\
{\mathcal D}_0^3 R = -\frac{1}{2} Q_1 +2y_0S_0 -\frac{1}{2}y_0^2 Q_0 -\frac{1}{4}Q_0^2-2P_0.
\label{RAzerothree.b}
\end{gather}
\end{subequations}
Substituting \eqref{RAeval}, \eqref{DDzeroR}, \eqref{RAzerothree.b},  into \eqref{ASVAiryR} and
using \eqref{xzeroAiry} shows that all terms cancel, verifying the equation.

\subsection{The Bessel kernel system}

 In this case, the system of dynamical equations is given \cite{TW3} by
\begin{subequations}
\label{BesselHameq}
\begin{gather}
\frac{\partial x_j}{\partial a_k} =-\frac{1}{2} \frac{(x_j y_k-y_jx_k)x_k}{a_j-a_k},
\qquad  j\neq k, \label{BesselHameqa.a}\\
\frac{\partial y_j}{\partial a_k} =-\frac{1}{2} \frac{(x_j y_k-y_jx_k)y_k}{a_j-a_k},
 \qquad j\neq k, \label{BesselHameqb.b}\\
a_j\frac{\partial x_j}{\partial a_j} =\frac{1}{2}\sum_{k=1\atop k\neq j}^n
\frac{a_k(x_j y_k-y_jx_k)x_k}{a_j-a_k} +2y_j +\frac{1}{4}y_0x_j,
\label{BesselHameqc.c}\\
a_j\frac{\partial y_j}{\partial a_j} =\frac{1}{2}\sum_{k=1\atop k\neq j}^n
\frac{a_k(x_j y_k-y_jx_k)y_k}{a_j-a_k}+\frac{1}{8}(\alpha^2-a_j+x_0)x_j
 -\frac{1}{4} y_0 y_j, \label{BesselHameqd.d}\\
\frac{\partial x_0}{\partial a_j}= -x_j y_j,  \qquad
\frac{\partial y_0}{\partial a_j} =-\frac{1}{4}x_j^2. \label{BesselHameqe.e}
\end{gather}
\end{subequations}
This is again a compatible system of nonautonomous Hamiltonian equations
generated by the Poisson commuting Hamiltonians $a_jG_{\alpha,j}^B$ defined
in~\eqref{GBjdef}, provided the Poisson brackets are defined by
\begin{equation}
\{x_j, \; y_k\}=\frac{1}{a_j}\delta_{jk}, \qquad \{x_0, \; y_0\}=-4 .
\end{equation}

  There again exist two additional conserved quantities for this case. The first
is defined by
\begin{equation}
G_0^B:=x_0 + \frac{1}{4}y_0^2 + y_0 + \frac{1}{4}Q_1,  \label{GBzero}
\end{equation}
as may be seen directly by differentiating with respect to the $a_j$'s, using
\eqref{BesselHameq}.  Since all the quantities appearing in~\eqref{GBzero} vanish
in the limit $\{a_j \to 0, \ \forall\; j\}$, this difference must vanish, and therefore
the invariant relation
\begin{equation}
x_0=-\frac{1}{4}y_0^2 - y_0 - \frac{1}{4}Q_1  \label{xzeroBessel}
\end{equation}
is satisfied for this solution. The second conserved quantity is
\begin{gather}
G^B_\infty := y_0 + 4 \sum_{j=1}^{2n} a_j G^B_{\alpha,j} = y_0 +4R^B_{\alpha,1} \nonumber\\
\phantom{G^B_\infty :}{} = y_0 -\frac{1}{4}\left(\alpha^2+x_0\right)Q_0+
\frac{1}{4}Q_1 +y_0 S_0 + 4P_0 + Q_0 P_0  -
S_0^2, \label{GBinfty}
\end{gather}
Again, due to the Poisson commutativity of the Hamiltonians defined in
\eqref{GBjdef}, the Hamiltonian vector fields generating the $a_j$ deformations when applied
to the term $R^B_{\alpha,1}$ give zero,  and therefore only the explicit dependence of this
term upon the parameters need be taken into account when verifying that differentiation
of the sum gives zero. Since all the quantities appearing in~\eqref{GBinfty} vanish in the
limit $\{a_j\to 0, \ \forall\; j\}$, the invariant $G^B_\infty$ must also vanish on this
particular solution, and we therefore have the relation
\begin{gather}
y_0 = -4R^B_{\alpha,1}=-4{\mathcal D}_1F^B_{\alpha}\nonumber\\
\phantom{y_0}{}=\frac{1}{4} \left(\alpha^2 +x_0\right)Q_0
-\frac{1}{4}Q_1 -y_0 S_0-4P_0 -Q_0 P_0 +S_0^2. \label{yzeroBessel}
\end{gather}

The quantities $R^B_{\alpha,1}$, $R^B_{\alpha,2}$ are given by
\begin{subequations}
\label{DDoneRB}
\begin{gather}
R^B_{\alpha,1}= {\mathcal D}_1 F^B_{\alpha}= \sum_{j=1}^{2n}a_j G^B_{\alpha,j}  \nonumber\\
\phantom{R^B_{\alpha,1}}{} = -\frac{1}{16}\left(\alpha^2 + x_0\right)Q_0 +\frac{1}{16}Q_1
 + \frac{1}{4}y_0S_0 + P_0  + \frac{1}{4} Q_0 P_0 - \frac{1}{4}S_0^2,
\label{DDoneRB.a}\\
R^B_{\alpha,2}= {\mathcal D}_2 F^B_{\alpha}= \sum_{j=1}^{2n}a_j^2 G^B_{\alpha,j} \nonumber\\
\phantom{R^B_{\alpha,2}}{}=  -\frac{1}{16}\left(\alpha^2+ x_0\right)Q_1 +\frac{1}{16}Q_2
 + \frac{1}{4}y_0S_1  + P_1.  \label{DDtwoRB.b}
\end{gather}
\end{subequations}
It again follows from the Poisson commutativity of the Hamiltonians $\{G^B_{\alpha,j}\}$
that the derivatives of $R^B_{\alpha,1}$ and  $R^B_{\alpha,2}$ with respect to the parameters
are given  by their explicit dependence on these parameters, and hence
\begin{subequations}
\label{DDoneRone}
\begin{gather}
{\mathcal D}_1^2F^B_\alpha = {\mathcal D}_1 R^B_{\alpha,1}= \frac{1}{16}Q_1, \label{DDoneRone.a}\\
{\mathcal D}_2{\mathcal D}_1 F^B_\alpha= {\mathcal D}_2 R^B_{\alpha,1}
= \frac{1}{16}Q_2.  \label{DDtwoRone.b}
\end{gather}
\end{subequations}

From \eqref{BesselHameq}, application of ${\mathcal D}_1$ to
$\{Q_1,S_1,x_0, y_0\}$  gives
\begin{subequations}
\label{QSBonederiv}
\begin{gather}
{\mathcal D}_1 Q_1 =Q_1 + 4S_1 +\frac{1}{2}y_0 Q_1, \qquad
{\mathcal D}_1 S_1 = S_1 + \frac{1}{8}\left(\alpha^2 +x_0\right)Q_1 -\frac{1}{8}Q_2  + 2P_1,
\label{QSBonederiv.a}\\
{\mathcal D}_1 x_0 =  -S_1, \qquad {\mathcal D}_1 y_0 =-\frac{1}{4}Q_1.  \label{xyBzeroderiv.b}
\end{gather}
\end{subequations}
 Further application of ${\mathcal D}_1$, using \eqref{DDzeroR.a}, \eqref{QSBonederiv}, and
\eqref{xzeroBessel} therefore gives
\begin{subequations}
\label{DDthree}
\begin{gather}
{\mathcal D}_1^3 F^B_\alpha = \frac{1}{16}\left(1 + \frac{y_0}{2}\right)Q_1 +\frac{1}{4} S_1,
\label{DDthree.a}\\
{\mathcal D}_1^4 F^B_\alpha = \frac{1}{16}\left(1 + \frac{\alpha^2}{2} +\frac{y_0}{2}
+ \frac{y_0^2}{8}\right)Q_1\nonumber\\
\phantom{{\mathcal D}_1^4 F^B_\alpha =}{}+\frac{1}{2}P_1 + \left(\frac{1}{2} +\frac{y_0}{8}\right) S_1  - \frac{1}{64}Q_1^2
-\frac{1}{32} Q_2.
\label{DDfour.b}
\end{gather}
\end{subequations}
Substitution of \eqref{DDtwoRB.b}, \eqref{DDoneRone}, \eqref{DDthree} into
\eqref{ASVBessel}, and use of  \eqref{yzeroBessel} to replace the term $-4{\mathcal D}_1
F^B_\alpha$ by
$y_0$,  and \eqref{xzeroBessel} to eliminate $x_0$, shows that all the terms cancel,
verifying the equation.

\section{Deduction of the $\boldsymbol{\tau}$-function  equations\\
from isomonodromic deformations}

In this section, we show how the full hierarchies of equations derived in
\cite{ASV1, ASV2} may be deduced from the Hamiltonian systems \eqref{AiryHameq},
\eqref{BesselHameq} and also how the corresponding hierarchy is deduced for
the case of the sine kernel. The key step is to recast these systems as isomonodromic
deformation equations for an associated differential operator in an auxiliary spectral
variable
$z\in {\mathbb P}^1$, having rational coefficients with poles at the points $\{z=a_j\}$, and
to interpret the Fredholm determinants
$\tau^S$,
$\tau^A$ and $\tau^B_\alpha$ as isomonodromic $\tau$-functions.

\subsection{The Airy kernel isomonodromic system}

 The Hamiltonian system \eqref{AiryHameq} implies that the compatibility
conditions
\begin{subequations}
\label{AjkAiry}
\begin{gather}
\frac{\partial A_j}{\partial a_k} = \frac{[A_j, A_k]}{a_j -a_k} , \qquad j\neq k,  \label{AjkAiry.a}\\
\frac{\partial A_j}{\partial a_j} = [a_j B +C, A_j]
- \sum_{k=1\atop k\neq j}^{2n}\frac{[A_j, A_k]}{a_j -a_k},
\label{AjjAiry.b}\\
\frac{\partial C}{\partial a_j} = [B, A_j] \label{CkAiry.c}
\end{gather}
\end{subequations}
are satisfied for the following overdetermined system~\cite{HTW}
\begin{subequations}
\label{PsiAz}
\begin{gather}
\frac{\partial \Psi^A}{\partial z} = X^A(z)\Psi^A,  \label{PsiAz.a}\\
\frac{\partial \Psi^A}{\partial a_j} = -\frac{A_j}{z- a_j}\Psi^A, \qquad j=1, \dots 2n,
\label{PsiAaj.b}\\
 X^A(z)  := zB + C + \sum_{j=1}^{2n} \frac{A_j}{z-a_j}, \label{XAdef.c}
\end{gather}
\end{subequations}
where $\Psi^A(z, a_1, \dots a_{2n})$ is a $2\times 2$ matrix, invertible
where defined, and
\begin{subequations}
\label{Ajdef}
\begin{gather}
A_j  := -\frac{1}{2} \left(\begin{matrix} x_j y_j & y_j^2 \\ -x_j^2 & -x_j y_j\end{matrix}\right),
\label{Ajdef.a}\\
B  :=  \left(\begin{matrix} 0 & -\frac{1}{2} \\ 0 & 0\end{matrix}\right),
\qquad
C  := \left(\begin{matrix} y_0 & \frac{x_0}{2} \\ -2 & - y_0\end{matrix}\right). \label{BCAirydef.b}
\end{gather}
\end{subequations}
This implies the invariance of the monodromy of the operator $\frac{\partial}{\partial z} -
X^A(z)$ under changes in the parameters $\{a_j\}$. In view of eq.~\eqref{GAjdef}, according
to the constructions of \cite{JMU, JM}, the Fredholm determinant $\tau^A$ is just the
isomonodromic $\tau$-function of the system \eqref{AjkAiry}--\eqref{PsiAz}.

Now define the sequence of $2\times 2$ matrices
\begin{equation}
B_m := \sum_{j=1}^{2n} a_j^mA_j = -\frac{1}{2} \left(\begin{matrix} S_m & P_m \\ -Q_m & -S_m
\end{matrix}\right),
\qquad m\in {\mathbb N}, \label{Bmdef}
\end{equation}
where the quantities $P_m$, $Q_m$, $S_m$ were defined in~\eqref{PQS}.
Expanding $X^A(z)$ for large $z$ gives
\begin{equation}
X^A(z)= zB + C + \sum_{m=0}^\infty \frac{B_m}{z^{m+1}}.  \label{XAexp}
\end{equation}
Since
\begin{equation}
G^A_j = \frac{1}{2}\,\mbox{res}_{z=a_j} \, \mbox{tr}\left(\left(X^A\right)^2(z)\right),
\end{equation}
and
\begin{equation}
G^A_0 = \frac{1}{2}\,\mbox{res}_{z=\infty} \frac{1}{z}\,
 \mbox{tr}\left(\left(X^A\right)^2(z)\right),
\end{equation}
we have
\begin{equation}
\frac{1}{2}\,\mbox{tr}\left(\left(X^A\right)^2(z)\right)= z + G^A_0 +
\sum_{m=0}^{\infty} \frac{R^A_m}{z^{m+1}},
\end{equation}
where
\begin{equation}
R^A_m:= \sum_{j=1}^{2n}a_j^mG^A_j =  \mbox{tr}\,(B B_{m+1} + CB_m) + \frac{1}{2}\,
\mbox{tr}\sum_{k=0}^{m-1} B_k B_{m-k-1}
\label{RAmexp}
\end{equation}
(with the last term absent if $m=0$) are the quantities defined in~\eqref{RAmdef}.

  Using the fact that the Hamiltonian vector fields generating the $a_j$ deformations
give zero when applied to the $G^A_j$'s, and hence also the $R^A_m$'s, it follows that
the effect of applying the operators ${\mathcal D}_k$ to  $R^A_m$ gives just
the explicit derivatives,
\begin{equation}
{\mathcal D}_k R^A_m = (m+1)\,\mbox{tr}\, (B B_{m+k}) + m\,\mbox{tr}\, (CB_{m+k-1}) +
\sum_{l=1}^{m-1} l \,\mbox{tr}\,(B_{l+k-1}B_{m-l-1})  \label{DkRAm}
\end{equation}
(with the sum in the last term absent if $m=0$ and the second term absent if $m+k=0$).

  Applying the operator ${\mathcal D}_m$ to $\Psi^A$, using \eqref{PsiAaj.b} and
\eqref{Bmdef} gives the
sequence of equations
\begin{equation}
{\mathcal D}_m\Psi^A = -\sum_{k=0}^\infty
\frac{B_{m+k}}{z^{k+1}}\Psi^A, \qquad m\in {\mathbb N}.
\end{equation}
The compatibility of these equations with \eqref{PsiAz.a} implies the following equations
for the matrices $\{B_m, C\}$.
\begin{subequations}
\label{DkBAm}
\begin{gather}
{\mathcal D}_k B_m  = m B_{m+k-1} + [C, B_{m+k}] + [B, B_{m + k + 1}]
+ \sum_{l=0}^{m-1}[B_l, B_{m+k-l-1}], \label{DkBAm.a}\\
{\mathcal D}_k C  = [B, B_k], \qquad  k,m \in {\mathbb N}   \label{DkCAm.b}
\end{gather}
\end{subequations}
(where the first term of \eqref{DkBAm.a} is absent if $m+k=0$ and the last
term is absent if $m=0$).

   The strategy for deriving the hierarchy of equations for $\tau^A$ is to now choose
a~$k$-va\-lue $(k_1)$ in \eqref{DkRAm}, \eqref{DkBAm}  and use
these equations, together with \eqref{RAmexp} to express all the
relevant matrix elements of the $B_m$'s  for $m\le k$ in terms of
the $R_{k}$'s  for  $k < k_1$ and the corresponding ${\mathcal
D}_{k}$'s applied repeatedly to them. Equations  \eqref{DkBAm},
for $k=k_1$ may then be expressed entirely in terms of these
quantities, and hence in terms of repeated applications of the
operators ${\mathcal D}_{k}$ to $F^A=\ln \tau^A$. An essential
step in this procedure is to also eliminate the additional
variables $x_0$, $y_0$ from the equations through use of the
invariant conditions~\eqref{xzeroAiry}, \eqref{yzeroAiry}.

   For example, choosing $k_1 = 1 $, we note that  for $m=0$, eq.~\eqref{RAmexp}
reduces to \eqref{RAeval} while
for $k=0,1$ and $m=0$, \eqref{DkRAm} reduces to \eqref{DDzeroR}
and for $k=0, m=0$, eqs.~\eqref{DkBAm}
give \eqref{PSzeroAderiv.a}, \eqref{xyAzeroderiv.b}. Combining
these with the invariant relations \eqref{xzeroAiry}, \eqref{yzeroAiry} allows us to express the
relevant matrix elements of $C$, $B_0$ and $B_1$ as
\begin{subequations}
\begin{gather}
x_0 = {\mathcal D}_0 R +R^2, \qquad
y_0=R, \\
Q_0 =-4{\mathcal D}_0R, \qquad  S_0 = -2R{\mathcal D}_0 R - {\mathcal D}_0^2 R, \\
P_0 = \frac{1}{2}R -\frac{1}{4}{\mathcal D}_0^3R -R{\mathcal D}_0^2 R -
\frac{1}{2}({\mathcal D}_0R)^2
-R^2 {\mathcal D}_0 R,\\
Q_1 =-2R-6({\mathcal D}_0R)^2-{\mathcal D}_0^3 R.
\end{gather}
\end{subequations}
Substituting these in eq.~\eqref{DkCAm.b} for $k=1$ gives
\eqref{ASVAiryR}. Similarly, eq.~\eqref{DkBAm.a} for $k=1$, $m=0$
and eq.~\eqref{RAmexp} for $m=1$ produce the following expressions
for the relevant matrix elements of $B_1$ and $B_2$.
\begin{subequations}
\label{AirySone}
\begin{gather}
S_1 =-{\mathcal D}_1{\mathcal D}_0R  -R^2 - 3R({\mathcal D}_0R)^2 - R{\mathcal D}_0^3R,
 \label{AirySone.a}\\
Q_2 =  -2R_1 -{\mathcal D}_1{\mathcal D}_0^2 R -2({\mathcal D}_0 R)({\mathcal D}_1 R)
-R{\mathcal D}_0R -\frac{3}{2}R{\mathcal D}_1{\mathcal D}_0R  - \frac{3}{2}
({\mathcal D}_0R)\left({\mathcal D}_0^3\right)R \nonumber\\
\phantom{Q_2=}{} +\frac{1}{2}\left({\mathcal D}_0^2
R\right)^2 -\frac{3}{2}R^3 -\frac{1}{2}R^2{\mathcal D}_0^3R -
 7({\mathcal D}_0 R)^3  -\frac{9}{2}R^2 ({\mathcal D}_0R)^2.
\label{AiryQtwo.b}
\end{gather}
\end{subequations}
Substitution of \eqref{AiryQtwo.b} in eq.~\eqref{DkBAm.a} (or \eqref{DkRAm})
for $k=2, m=0$,  thus gives
\begin{gather}
4{\mathcal D}_1R  - 2R_1 -{\mathcal D}_1{\mathcal D}_0^2 R -
2({\mathcal D}_0 R)({\mathcal D}_1 R)
-R{\mathcal D}_0R -\frac{3}{2}R{\mathcal D}_1{\mathcal D}_0R  -
\frac{3}{2}({\mathcal D}_0R)\left({\mathcal D}_0^3\right)R \nonumber\\
\phantom{4{\mathcal D}_1R} {}+\frac{1}{2}\left({\mathcal D}_0^2
R\right)^2 -\frac{3}{2}R^3 -\frac{1}{2}R^2{\mathcal D}_0^3R -
7({\mathcal D}_0 R)^3  -\frac{9}{2}R^2 ({\mathcal D}_0R)^2 = 0.
\end{gather}
as the next equation of the hierarchy. The remaining equations may
similarly be expressed in terms of the derivations ${\mathcal D}_{k}$ acting upon $F^A$.

\subsection{The Bessel kernel isomonodromic system}

{\samepage
The Bessel kernel case is so similar to the above that only the pertinent equations
will be given, without repeating any details of the procedure.
Define for this case, the matrices
\begin{subequations}
\label{XBdef}
\begin{gather}
X^B(z)  := \widetilde{B} + \frac{C_\alpha -\sum\limits_{j=1}^{2n}A_j}{z} + \sum_{j=1}^{2n}
\frac{A_j}{z-a_j},\label{XBdef.a}\\
 \widetilde{B}  :=  \left(\begin{matrix} 0 & \frac{1}{8} \\ 0 & 0\end{matrix}\right), \qquad
C_\alpha := -\frac{1}{4} \left(\begin{matrix} y_0 & \frac{1}{2}\left(x_0 +\alpha^2\right) \\
 8 & - y_0\end{matrix}\right).
\label{XBdef.b}
\end{gather}
\end{subequations}
where the $A_j$'s are again defined as in~\eqref{Ajdef.a}.}

The Hamiltonian system \eqref{BesselHameq} implies that the compatibility
conditions
\begin{subequations}
\label{AjkBessel}
\begin{gather}
\frac{\partial A_j}{\partial a_k} = \frac{[A_j, A_k]}{a_j -a_k} , \qquad j\neq k,
\label{AjkBessel.a}\\
a_j\frac{\partial A_j}{\partial a_j} = [C_\alpha +a_j \widetilde{B}, A_j]
- \sum_{k=1\atop k\neq j}^{2n}\frac{a_k[A_j, A_k]}{a_j -a_k},  \label{AjjBessel.b}\\
\frac{\partial C_\alpha}{\partial a_j} = [\widetilde{B}, A_j] \label{CkBessel.c}
\end{gather}
\end{subequations}
are satisfied  for the system
\begin{subequations}
\label{PsiBz}
\begin{gather}
\frac{\partial \Psi^B}{\partial z} = X^B(z)\Psi^B,  \label{PsiBz.a}\\
\frac{\partial \Psi^B}{\partial a_j} = -\frac{A_j}{z- a_j}\Psi^B, \qquad j=1, \dots 2n,
\label{PsiBaj.b}
\end{gather}
\end{subequations}
where $\Psi^B(z, a_1, \ldots, a_{2n})$ is again a $2\times 2$ matrix, invertible
where defined. This again implies the invariance of the monodromy of the operator
$\frac{\partial}{\partial z} - X^B(z)$ under changes in the parameters $\{a_j\}$.
In view of eq.~\eqref{GBjdef}, the Fredholm determinant $\tau^B_\alpha$ is again an
isomonodromic $\tau$-function for the system \eqref{AjkBessel}--\eqref{PsiBz}.

Defining the sequence of $2\times 2$ matrices $\{B_m ,\ m\in {\mathbb N}\}$  as in
\eqref{Bmdef}, and expanding $X^B(z)$ for large $z$ gives
\begin{equation}
X^B(z)= \widetilde{B} + \frac{C_\alpha}{z} + \sum_{m=1}^\infty \frac{B_m}{z^{m+1}},  \label{XBexp}
\end{equation}
and
\begin{equation}
\frac{1}{2}\,\mbox{tr}\left(\left(X^B\right)^2(z)\right)=-\frac{1}{4} z +
\frac{G^B_0-G^B_\infty+\alpha^2}{4z^2} +
\sum_{m=1}^{\infty} \frac{R^B_{\alpha,m}}{z^{m+1}},
\end{equation}
where
\begin{subequations}
\label{RBoneexp}
\begin{gather}
R^B_{\alpha,1}=\frac{1}{4}\left(G^B_\infty-G^B_0 -\alpha^2\right)+ \frac{1}{2}\,
\mbox{tr}\left(C_\alpha^2
+2\widetilde{B} B_1\right), \label{RBoneexp.a}\\
R^B_{\alpha,m} =  \widetilde{tr}\left(\widetilde{B} B_{m} + C_\alpha B_{m-1}\right)
 + \frac{1}{2}\,\mbox{tr}\sum_{k=1}^{m-2} B_k B_{m-k-1}, \qquad m\ge 2
\label{RBmexp.b}
\end{gather}
\end{subequations}
are the quantities defined in~\eqref{RBmdef} and $G_0^B$,
$G_\infty^B$ are the conserved quatities defined
in~\eqref{GBzero}, \eqref{GBinfty}, which vanish on the particular
solutions defined by \eqref{xjdef}.

 The fact that the Hamiltonian vector fields generating the $a_j$ deformations
give zero when applied to the $G^B_{\alpha,j}$'s, and  $R^B_{\alpha,m}$'s again implies that
the effect of applying the operators ${\mathcal D}_k$ to the $R^B_{\alpha,m}$'s is to evaluate
only explicit derivatives with respect to the parameters, giving
\begin{gather}
{\mathcal D}_k R^B_{\alpha,1} = \frac{1}{2} \,\mbox{tr}\left(\widetilde{B} B_k\right),\nonumber\\
{\mathcal D}_k R^B_{\alpha,m} = m\,\mbox{tr}\left(\widetilde{B} B_{m+k-1}\right)
+ (m-1) \,\mbox{tr}\, (C_\alpha B_{m+k-2})\nonumber\\
\phantom{{\mathcal D}_k R^B_{\alpha,m} =}{} +\sum_{l=1}^{m-2} l\, \mbox{tr}(B_{l+k-1}B_{m-l-1}), \qquad m\ge 2 \label{DkRBm}
\end{gather}
(with the sum in the last term absent if $m=2$).

  Applying the operator ${\mathcal D}_m$ to $\Psi^B$,
using \eqref{Bmdef} and \eqref{PsiBaj.b}, again gives the
sequence of equations
\begin{equation}
{\mathcal D}_m\Psi^B = -\sum_{k=0}^\infty
\frac{B_{m+k}}{z^{k+1}}\Psi^B, \qquad m\in {\mathbb N}.
\end{equation}
whose compatibility  with \eqref{PsiBz.a} implies the following equations
for the matrices $\{B_m, C_\alpha\!\},$
\begin{subequations}
\label{DkBBm}
\begin{gather}
{\mathcal D}_k B_m  = m B_{m+k-1} + [C_\alpha, B_{m+k-1}] + [\widetilde{B}, B_{m+k}]
+ \sum_{l=1}^{m-1}[B_l, B_{m+k-l-1}], \label{DkBBm.a}\\
{\mathcal D}_k C_\alpha  = [\widetilde{B},B_k], \qquad  k,m \in {\mathbb N}, \quad m\ge 1.
  \label{DkCBm.b}
\end{gather}
\end{subequations}

   The hierarchy of equations for $\tau^B_\alpha$ is derived in the same way as for
the Airy case.  For example,  eqs.~\eqref{RBoneexp} for $k=2$
reduce to \eqref{DDoneRB}, while \eqref{DkRBm} for $k=1,2$, $m=1$
reduces to~\eqref{DDoneRone}, and eqs.~\eqref{DkBBm} for $k=1,2$,
$m=1$ give \eqref{QSBonederiv}. Combining these with the invariant
relations \eqref{xzeroBessel}, \eqref{yzeroBessel}
 allows us to
express the relevant matrix elements of $C_\alpha$, $B_1$
and~$B_2$ as
\begin{subequations}
\begin{gather}
x_0 = -4\left({\mathcal D}_1 R^B_{\alpha,1} + \left( R^B_{\alpha,1}\right)^2 - 4R^B_{\alpha,1}\right),
\qquad  y_0 = -4R^B_{\alpha,1},  \\
Q_1 =16{\mathcal D}_1 R^B_{\alpha,1}, \\
S_1 = 8R^B_{\alpha,1}{\mathcal D}_1 R^B_{\alpha,1} -4 {\mathcal D}_1R^B_{\alpha,1}
+ 4{\mathcal D}^2_1 R^B_{\alpha,1}, \\
 P_1 =  R^B_{\alpha,2} +\alpha^2R^B_{\alpha,1} + 4\left(R^B_{\alpha,1}\right)^2
 {\mathcal D}_1R^B_{\alpha,1} -4\left({\mathcal D}_1 R^B_{\alpha,1}\right)^2 \nonumber\\
\phantom{P_1 =}{}+ 4 R^B_{\alpha,1} {\mathcal D}^2_1 R^B_{\alpha,1} - {\mathcal D}_2 R^B_{\alpha,1}, \\
 Q_2 =16{\mathcal D}_2 R^B_{\alpha,1}.
\end{gather}
\end{subequations}
Substituting these in eqs.~\eqref{DkBBm} for $k=2$ gives \eqref{ASVBessel}. Similar
calculations for higher values of $k$ yield the further equations of the
Bessel hierachy.

\subsection{The sine kernel system}

For this case, the quantities  defined in \eqref{xjdef.a}--\eqref{yjdef.b} satisfy the system of
dynamical equations defined in \cite{JMMS, TW1}
\begin{subequations}
\label{SineHameqa}
\begin{gather}
\frac{\partial x_j}{\partial a_k} =-\frac{1}{2} \frac{(x_j y_k-y_jx_k)x_k}{a_j-a_k},
\qquad  j\neq k, \label{SineHameqa.a}\\
\frac{\partial y_j}{\partial a_k} =-\frac{1}{2} \frac{(x_j y_k-y_jx_k)y_k}{a_j-a_k},
 \label{SineHameqb.b}
\end{gather}
\begin{gather}
\frac{\partial x_j}{\partial a_j} =\frac{1}{2}\sum_{k=1\atop k\neq j}^n
\frac{(x_j y_k-y_jx_k)x_k}{a_j-a_k} +2y_j,  \label{SineHameqc.c}
\\
\frac{\partial y_j}{\partial a_j} =\frac{1}{2}\sum_{k=1\atop k\neq j}^n
\frac{(x_j y_k-y_jx_k)y_k}{a_j-a_k}-\frac{\pi^2}{2}x_j.
\label{SineHameqd.d}
\end{gather}
\end{subequations}
This is again a  compatible system of nonautonomous Hamiltonian
equations, generated by the Poisson commuting Hamiltonians
$\{G^S_j\}$ defined in~\eqref{GSjdef}.  They imply that the
compatibility conditions
\begin{subequations}
\label{Ajksine}
\begin{gather}
\frac{\partial A_j}{\partial a_k} = \frac{[A_j, A_k]}{a_j -a_k} ,
\qquad j\neq k, \label{Ajksine.a}\\ \frac{\partial A_j}{\partial
a_j} = [B_S, A_j] - \sum_{k=1\atop k\neq j}^{2n}\frac{[A_j,
A_k]}{a_j -a_k}, \qquad j\neq k \label{Ajjsine.b}
\end{gather}
\end{subequations}
are satisfied for the system
\begin{subequations}
\label{PsiSz} \begin{gather} \frac{\partial \Psi^S}{\partial z} =
X^S(z)\Psi^S, \label{PsiSz.a}\\ \frac{\partial \Psi^S}{\partial
a_j} = -\frac{A_j}{z- a_j}\Psi^S, \qquad j=1, \dots, 2n,
\label{PsiSaj.b} \end{gather} \end{subequations} where
\begin{subequations}
\label{XSdef} \begin{gather}
 X^S(z)  := B_S + \sum_{j=1}^{2n}
\frac{A_j}{z-a_j}, \label{XSdef.a}\\
 B_S := \left(\begin{matrix} 0 &
\frac{\pi^2}{2} \\ -2 & 0\end{matrix}\right), \label{XSdef.b}
\end{gather}
\end{subequations}
 with the $A_j$'s again defined as in~\eqref{Bmdef}. As in the
previous cases, this implies the invariance of the monodromy of
the operator $\frac{\partial}{\partial z} - X^S(z)$. In view of
eq.~\eqref{GSjdef}, the Fredholm determinant $\tau^S$ is an
isomonodromic $\tau$-function for the system
\eqref{Ajksine}--\eqref{PsiSz}.

Expanding $X^S(z)$ for large $z$ gives
\begin{equation}
X^S(z)= B_S + \sum_{m=0}^\infty \frac{B_m}{z^{m+1}}, \label{XSexp}
\end{equation}
with the matrices $\{B_m ,\ m\in {\mathbb N}\}$ again defined as
in~\eqref{Bmdef}, and \begin{equation}
\frac{1}{2}\,\mbox{tr}\left(\left(X^S\right)^2(z)\right)=-\pi^2 +
\sum_{m=0}^{\infty} \frac{R^S_m}{z^{m+1}}, \end{equation}
 where
 \begin{equation}
R^S_m := \sum_{j=1}^{2n}a_j^m G^S_j = \mbox{tr}\,(B_S B_m)
 + \frac{1}{2}\,\mbox{tr}\sum_{k=0}^{m-1} B_k B_{m-k-1}, \qquad
 m\in {\mathbb N}. \label{RSmexp}
\end{equation}

 Applying the operators ${\mathcal D}_k$ to $R^S_m$ again just differentiates
explicitly with respect to the parameters, giving
\begin{equation}
{\mathcal D}_k R^S_m = m\,\mbox{tr}\, (B_S B_{m+k-1}) +
\sum_{l=1}^{m-1} l\, \mbox{tr}\,(B_{l+k-1}B_{m-l-1}) \label{DkRSm}
\end{equation}
 (with the first term absent if $k + m=0$ and the sum in
the last term absent if $m=0$).

  Applying ${\mathcal D}_m$ to $\Psi^S$, using \eqref{XSexp}
  and \eqref{PsiSaj.b},  gives the
sequence of equations
\begin{equation}
{\mathcal D}_m\Psi^S = -\sum_{k=0}^\infty
\frac{B_{m+k}}{z^{k+1}}\Psi^S, \qquad m\in {\mathbb N},
\end{equation}
 whose compatibility with~\eqref{PsiSz.a} implies the following
equations for the matrices $\{B_m\}$, \begin{equation}
 {\mathcal
D}_k B_m = m B_{m+k-1} + [B_S, B_{m+k}] + \sum_{l=0}^{m-1}[B_l,
B_{m+k-l-1}]. \label{DkBSm}
\end{equation}

   The hierarchy of equations for $\tau^S$ is derived in the same way as for
the Airy and Bessel cases, except that we no longer have two
conserved quantities like $G^{A,B}_0$, $G^{A,B}_\infty$. To derive
a closed system of equations, we are obliged to include two
further dependent variables~$\tau^S_{\pm}$, which we choose as the
nonvanishing entries of the matrix $[B_S,B_0]\tau^S$,
\begin{equation}
\tau_+^S:= \left(2P_0 -\frac{\pi^2}{2}Q_0\right)\tau^S, \qquad
\tau_-^S:=S_0\tau^S. \label{tauSpmdef} \end{equation}
 The
remaining component of $B_0$, which cancels in the commutator
$[B_S,B_0]$, is \begin{equation}  R^S_0=\mbox{tr}\,(B_S B_0) = P_0
+\frac{\pi^2}{4}Q_0 = 0, \label{RSzero} \end{equation}
 where the
first equality follows from choosing $m=0$ in~\eqref{RSmexp}. This
provides a single conserved quantity that vanishes for the
particular solution defined by \eqref{xjdef.a}--\eqref{yjdef.b}.

  To derive the hierarchy of $\tau$-function equations, we first combine
eqs.~\eqref{tauSpmdef}--\eqref{RSzero}, which allows us to express
the matrix elements of $B_0$ as \begin{equation}  Q_0 =
-\frac{\tau^S_+}{\pi^2\tau^S}, \qquad P_0
=\frac{\tau^S_+}{4\tau^S} ,\qquad  S_0 = \frac{\tau^S_-}{\tau^S}.
\label{QPSzero} \end{equation}
 Eq.~\eqref{DkBSm} for $k=0$, $m=0$ gives
 \begin{equation}
{\mathcal D}_0 P_0=-\pi^2 S_0, \qquad {\mathcal D}_0Q_0= 4S_0,
\qquad {\mathcal D}_0S_0 =2P_0 -\frac{\pi^2}{2}Q_0,
\label{DzeroQPSzero} \end{equation}
 and substituting \eqref{RSzero},
\eqref{QPSzero} in \eqref{DzeroQPSzero} gives
\begin{subequations}
\label{tauSzero} \begin{gather} {\mathcal D}_0\tau^S =0,
\label{tauSzero.a}\\
 {\mathcal D}_0\tau_-^S = \tau_+^S, \qquad
{\mathcal D}_0\tau_+^S=-4\pi^2\tau_-^S. \label{taupmzero.b}
\end{gather}
\end{subequations}
These equations are  the lowest ones in the sine kernel hierarchy;
note that they are linear because of the vanishing of the
invariant $R^S_0$. To obtain higher, nonlinear equations, we first
note that eq.~\eqref{RSmexp} for $m=1$ gives \begin{equation}
R^S_1= P_1 +\frac{\pi^2}{4}Q_1
+\frac{1}{4}\left(S_0^2-Q_0P_0\right), \label{RSone}
\end{equation}
 while \eqref{DkRSm} for $k=0,1$, $m=1$ reduces to
\begin{subequations}
\label{DDzeroRSone}
\begin{gather}
{\mathcal D}_0 R^S_1 = R_0=0, \label{DDzeroRSone.a}\\
 {\mathcal D}_1 R^S_1 = P_1 +\frac{\pi^2}{4}Q_1. \label{DDoneRSone.b}
 \end{gather}
 \end{subequations} The
first of these just gives the equation \begin{equation}
 {\mathcal D}_0{\mathcal D}_1\tau^S=0,
 \end{equation}
  which already follows from~\eqref{tauSzero.a}.
  The second, combined with eq.~\eqref{RSone} and eq.~\eqref{DkBSm} for
$k=1$, $m=0$ gives the further equation \begin{equation}
\tau^S{\mathcal D}^2_1\tau^S - \left({\mathcal
D}_1\tau^S\right)^2=\tau^S{\mathcal D}_1\tau^S -\frac{1}{4}
\left(\tau^S_-\right)^2 - \frac{1}{16
\pi^2}\left(\tau^S_+\right)^2 . \end{equation}
 Equation \eqref{DkBSm} for
$k=1$, $m=0$ gives \begin{equation}
 {\mathcal D}_1S_0 = 2P_1
-\frac{\pi^2}{2} Q_1, \qquad {\mathcal D}_1P_0 = -\pi^2 S_1,
\qquad {\mathcal D}_1Q_0 = 4S_1. \end{equation} Solving these,
together with \eqref{DDoneRSone.b}, gives the following
expressions for the matrix entries of~$B_1$:
\begin{subequations}
\label{QSone} \begin{gather}
 Q_1 = \frac{2}{\pi^2} \frac{{\mathcal
D}_1\tau^S}{\tau^S} -\frac{1}{2\pi^2}
\left(\frac{\tau^S_-}{\tau^S}\right)^2
-\frac{1}{8\pi^4}\left(\frac{\tau^S_+}{\tau^S}\right)^2 -
\frac{1}{\pi^2}{\mathcal D}_1\left(\frac{\tau^S_-}{\tau^S}\right),
\label{QSone.a}\\
 P_1 = \frac{1}{2} \frac{{\mathcal D}_1\tau^S}{\tau^S} -
 \frac{1}{8} \left(\frac{\tau^S_-}{\tau^S}\right)^2 -
 \frac{1}{32\pi^2}\left(\frac{\tau^S_+}{\tau^S}\right)^2
+\frac{1}{4}{\mathcal D}_1\left(\frac{\tau^S_-}{\tau^S}\right) ,
\label{PSone.b}\\
 S_1 = -\frac{1}{4\pi^2}{\mathcal D}_1\left(\frac{\tau^S_+}{\tau^S}\right).
  \label{SSone.c}
\end{gather}
\end{subequations}
Combining eq.~\eqref{DkBSm} for $(k=1,\; m=1)$ and for $(k=2,\;
m=0)$ gives \begin{equation}
 {\mathcal D}_2Q_0 = {\mathcal D}_1
Q_1 - Q_1, \qquad {\mathcal D}_2P_0 = {\mathcal D}_1 P_1 - P_1,
\qquad {\mathcal D}_2S_0 = {\mathcal D}_1 S_1 - S_1,
\label{DDtwo}\end{equation}
 Substitution  of \eqref{QPSzero}, \eqref{QSone} into
 \eqref{DDtwo} gives the next equations of the
hierarchy. Repeating this procedure for higher $(k, m)$ values
similarly generates the higher equations.

\section{Classical $\boldsymbol{R}$-matrix
approach and relation\\ to isospectral flows}

   In  \cite{ASV1, ASV2}, a key step in deriving the hierarchies of equations
for the Fredholm determinants~$\tau^A$ and $\tau^B_\alpha$ was to
begin with certain bilinear equations satisfied by KP
$\tau$-functions
 with respect to the flow
parameters $\{t_1, t_2, \ldots\}$ and to then use Virasoro
constraints to replace the $t_m$-derivations at vanishing
$t$-values by the operators ${\mathcal D}_m$. In this section, we
show how the classical $R$-matrix  approach to the underlying
isomonodromic deformation equations developed in \cite{H} provides
a direct link with commuting isospectral flows in the loop algebra
$\widetilde{\mathfrak{sl}}(2)$, without the requirement that these
arise as reduced KP flows. This fits into the broader framework of
commutative isospectral flows in loop algebras with respect to the
rational $R$-matrix Poisson  (or Adler--Kostant--Symes) structure
\cite{OPRS, AV, AHP, H} (and allows us to include the sine kernel
case, which does not appear as a reduced KP flow).

   First we recall \cite{H, HTW} that the isomonodromic deformation equations
\eqref{AjkAiry}, \eqref{AjkBessel}, \eqref{Ajksine} may be viewed
as Hamiltonian equations on the space of sets
$\{A_j\}_{j=1,\ldots, 2n}$ of $\mathfrak{sl}(2)$ elements, with
respect to the Lie Poisson bracket, extended in the Airy and
Bessel cases by the canonical variables $(x_0, y_0)$. (The
particular form \eqref{Ajdef.a} for the $A_j$'s  just represents
a~canonical parametrization on the symplectic leaves for which the
Casimir invariants $ \{\mbox{tr}\, A_j^2\}$ all vanish.) The
formulae \eqref{XAdef.c}, \eqref{XBdef.a}, \eqref{XSdef.a} define
a Poisson embedding of this space into the space
$\widetilde{\mathfrak{sl}}(2)^*_R$ of rational, traceless $2\times
2$ matrices depending rationally on the auxiliary loop variable
$z$, with respect to the Lie Poisson bracket on
$\widetilde{\mathfrak{sl}}(2)$ corresponding to the Lie bracket:
\begin{equation}
 [X,Y]_R :=\frac{1}{2}[RX,Y] + \frac{1}{2}[X, RY],
 \end{equation}
  where
\begin{equation}
 R :=P_+ -P_-
 \end{equation}
  is the {\it classical $R$-matrix},
given by the difference of the projection operators
\begin{gather}
P_+: \widetilde{\mathfrak{sl}}(2) \to
\widetilde{\mathfrak{sl}}_+(2), \qquad
P_+:\widetilde{\mathfrak{sl}}(2) \to
\widetilde{\mathfrak{sl}}_+(2),\nonumber\\
 P_-: X \to X_+, \qquad
P_-:X \to X_- \end{gather} to the subalgebras
$\widetilde{\mathfrak{sl}}_+(2)$, $\widetilde{\mathfrak{sl}}_-(2)$
consisting respectively of the nonnegative and negative terms in
the Laurent expanson of $X(z)$ for large~$z$. The space
$\widetilde{\mathfrak{sl}}(2)^*_R$ is identified as a~subspace of
$\widetilde{\mathfrak{sl}}(2)$ through the trace-residue pairing
\begin{equation}
\langle X, Y\rangle := {\mbox{res}}_{z=\infty}\,\mbox{tr}\,(X(z)
Y(z)).
\end{equation}

   In this setting, the isomonodromic deformation equations \eqref{AjkAiry},
\eqref{AjkBessel}, \eqref{Ajksine} may all be expressed in the
form \begin{subequations} \label{dXaj} \begin{gather}
 \frac{\partial X}{\partial a_j} = -[(dG_j)_-, X] + \frac{\partial (dG_j)_-}
 {\partial z}, \label{dXaj.a}\\
 (dG_j)_- =  - \frac{A_j}{z-a_j},  \label{dGjdef.b}
\end{gather}
\end{subequations}
where $X$ denotes $X^S$, $X^A$ or $X^B$, and $G_j$ denotes
$G_j^S$,  $G^A_j$ or $G^B_{\alpha,J}$ respectively. Viewing the
Hamiltonians $\{G_j\}$ as spectral invariants defined on the space
$\widetilde{\mathfrak{sl}}(2)$, eq.~\eqref{dXaj.a} follows from
the Adler--Kostant--Symes theorem, in view of the relations
\begin{equation}
\frac{\partial_0 X}{\partial a_j} = -\frac{\partial
(dG_j)_-}{\partial z}, \end{equation} where $\frac{\partial_0
X}{\partial a_j}$ denotes the derivative with respect to the {\it
explicit} dependence on the parameters~$\{a_j\}$ only.

   Rather than using the spectral invariants $\{G_j\}$
   as Hamiltonians, we consider
the Hamiltonian equations generated by the linear combinations
$R^S_m$, $R^A_m$ or $R^B_{\alpha,m}$ defined in \eqref{RSmdef},
\eqref{RAmdef}, \eqref{RBmdef}, which are all of the form
\begin{equation}
{\mathcal D}_m X = -[(dR_m)_-, X] + \frac{\partial
(dR_m)_-}{\partial z}, \label{DmXnonauton} \end{equation} with the
respective identifications for $X$ and $\{R_m\}$. These are just
equations \eqref{DkBAm}, \eqref{DkBBm} or \eqref{DkBSm}, depending
on the identification, since \begin{equation}
 R_m= \frac{1}{2}\,{\mbox{res}}_{z=\infty}z^m \,\mbox{tr}\, X^2(z),
 \end{equation} and therefore $dR_m$,
viewed as an element of $\widetilde{\mathfrak{sl}}(2)$, is just
\begin{equation}
dR_m = z^m X(z) = \sum_{k=0}^\infty \frac{B_{k}}{z^{k-m+1}}.
\label{dRm} \end{equation} implying \begin{equation}
 (dR_m)_-  = \sum_{k=0}^\infty \frac{B_{m+k}}{z^{k+1}}. \label{dRm+}
 \end{equation}

   If, instead of the nonautonomous systems occurring here because of the
identifications of the $a_j$'s as multi-time parameters, we
consider the autonomous systems generated by the {\it same} set of
Hamiltonians $\{R_0, R_1, \ldots\}$, denoting the corresponding
flow parameters $\{t_0, t_1, \ldots\}$, the resulting equations
have the isospectral form \begin{equation} \frac{\partial
X}{\partial t_m} =\pm [(dR_m)_{\pm}, X],  \label{DmXauton}
\end{equation} where either of the projections $(dR_m)_{\pm}$ may be
used, since the differential $dR_m$, given by~\eqref{dRm+},
commutes with~$X$. Although these systems are generated by the
same Hamiltonians as the nonautonomous systems
\eqref{DmXnonauton}, they of course do {\it not} generate
isomonodromic deformations of the operator
$\frac{\partial}{\partial z} - X(z)$, and in fact are not even
compatible with the systems \eqref{DmXnonauton}; however, they are
compatible amongst themselves, generating commuting isospectral
Hamiltonian flows. The close relationship between the autonomous
and associated nonautonomous systems implies a correspondence
between the structure of the resulting hierarchies.

To see this, we substitute the expressions  \eqref{XAdef.c},
\eqref{XBdef.a} and \eqref{XSdef.a} for $X(z)$ and \eqref{dRm+}
for $dR_m$ into \eqref{DmXauton} to obtain the systems
\begin{subequations}
\label{DkBAmataon} \begin{gather} \frac{\partial B_m}{\partial
t_m} = [C, B_{m+k}] + [B, B_{m + k +1}] + \sum_{l=0}^{m-1}[B_l,
B_{m+k-l-1}], \label{DkBAmataon.a}\\ \frac{\partial C}{\partial
t_m} = [B,B_k], \qquad  k,m \in {\mathbb N} \label{DkCAmauton.b}
\end{gather}
\end{subequations}
for $X=X^A$, \begin{subequations} \label{DkBBmauton}
\begin{gather} \frac{\partial B_m}{\partial t_m} = [C,
B_{m+k-1}] + [\widetilde{B}, B_{m+k}]   + \sum_{l=1}^{m-1}[B_l,
B_{m+k-l-1}], \label{DkBBmauton.a}\\ \frac{\partial
C_\alpha}{\partial t_m} = [\widetilde{B},B_k], \qquad  k,m \in
{\mathbb N}, \quad m\ge 1. \label{DkCBmauton.b} \end{gather}
\end{subequations}
for $X=X^B_\alpha$ and \begin{equation} \frac{\partial
B_m}{\partial t_m} = [B_S, B_{m+k}] + \sum_{l=0}^{m-1}[B_l,
B_{m+k-l-1}] \label{DkBSmauton} \end{equation} for $X=X^S$. These
only differ from the equations \eqref{DkBAm}, \eqref{DkBBm} and
\eqref{DkBSm} by the absence of the term $m B_{m+k-1}$ in the
right hand side of \eqref{DkCAmauton.b}, \eqref{DkCBmauton.b},
\eqref{DkBSmauton} and the replacement \begin{equation} {\mathcal
D}_m \to \frac{\partial}{\partial t_m} \end{equation} for the
derivation on the left hand side. The procedure for deriving
hierarchies for such systems is well known in the isospectral
context  (see, e.g.~\cite{FNR} for details); the recursive
procedure used in Section~3 above is just the analog of this
approach applied to the isomonodromic systems \eqref{DkBAm},
\eqref{DkBBm} and \eqref{DkBSm}.

   As a final point, it should be noted
   that almost nothing in the derivation of the
$\tau$-func\-tion equations of Sections~2 and~3 depended on the
fact that the specific $\tau$-functions involved were equal to the
Fredholm determinants \eqref{sineFredholmdet},
\eqref{AiryFredholmdet}, \eqref{BesselFredholmdet}. Everything
just followed from the general form of the isomonodromic
deformation equations \eqref{AjkAiry}, \eqref{AjkBessel} and
\eqref{Ajksine}, the only features specific to the identifications
of $\tau^A$, $\tau^B_\alpha$, $\tau^S$ as  Fredholm determinants
being the fact that the matrix residues $A_j$ were of rank~$1$ (as
seen from the parametrization~\eqref{Ajdef.a}) and the invariants
$G^A_0$, $G^A_\infty$, $G^B_0$,  $G^B_\infty$ vanished. By
allowing these invariants, as well as the constants $\{\det
A_j\}$, to take arbitrary values, an identical procedure leads to
equations for the $\tau$-functions of the general isomonodromic
systems, which only differ from the ones derived in Sections~2
and~3 by the nonzero constant values of the two additional
invariants $G^A_0$, $G^A_\infty$ or $G^B_0$ and $G^B_\infty$. For
example, eq.~\eqref{ASVAiryR} is replaced in the general case by
\begin{equation}
 {\mathcal D}_0^3 R -4{\mathcal D}_1 R + 2 R +
4\left(g_\infty^2-g_0\right){\mathcal D}_0R-2g_\infty
\left({\mathcal D}^2_0R +2R {\mathcal D}_0R\right) + 6 ({\mathcal
D}_0 R)^2 = 0, \label{ASVAiryRgeneral} \end{equation} where $g_0$,
$g_\infty$ are the values taken by the invariants $G_0^A$,
$G_\infty^A$, respectively. The other equations of these
hierarchies may similarly expressed in a way that allows arbitrary
values for these constants.

\subsection*{Acknowledgements} The author
would like to thank P. van  Moerbeke, C. Tracy and H. Widom for
helpful discussions relating to this work. Research supported in
part by the Natural Sciences and Engineering Research Council of
Canada and by the Fonds FCAR du Qu\'ebec.

\label{harnad-lastpage}

\end{document}